# Modelling the Semantic Web using a Type System

Rod Moten
PROARC, Inc.
Baltimore, MD
rmoten@proarc-inc.com

## ABSTRACT
We present an approach for modeling the Semantic Web as a type system. By using a type system, we can use symbolic representation for representing linked data. Objects with only data properties and references to external resources are represented as terms in the type system. Triples are represented symbolically using type constructors as the predicates. In our type system, we allow users to add analytics that utilize machine learning or knowledge discovery to perform inductive reasoning over data. These analytics can be used by the inference engine when performing reasoning to answer a query. Furthermore, our type system defines a means to resolve semantic heterogeneity on-the-fly.

## Categories and Subject Descriptors
D.3.3 [**Artificial Intelligence**]: Knowledge Representation Formalisms and Method – *representation languages, semantic networks.*

## General Terms
Languages.

## Keywords
Semantic Web, Linked Data, Type System

## 1. INTRODUCTION
The premier language for modelling the Semantic Web is RDF. RDF(S) follows an approach of modelling knowledge similar to description logics. This means inference is performed using statements that declare which classes individuals belong to (A-box statements) and statements that declare the relationships between classes and properties (T-Box statements). This mode of operation can be cumbersome when treating the Semantic Web as an open data space. By an open data space, we mean a distributed repository of data whereby any individual or organization can contribute to the data space. The only requirement is that the data be in a specific format, such as RDF, and retrievable by a specific protocol, such as HTTP. By following the description logic approach, if a new data provider wanted its data to be included in an existing query for data that has the same meaning, then the data provider would need to create the A-Box and T-Box statements so the query engine could deduce that the new data should be included in the search space of the query.

As an example, suppose there is an application that fuses together Person of Interest from a news organization and Birth Records from hospitals. Assume the Person of Interest has two attributes, name and date of birth. Let's call the Person of Interest from the news organization *A*. Assume the application was designed to retrieve Persons of Interest by querying the data space for *A*. Suppose after the application was implemented a new data provider was added to the system that also generates Persons of Interest. Ideally, the Persons of Interest from the new data provider, *B*, should also be included in the original query for Persons of Interest without changing the query. For this to happen, we have to explicitly state that each instance of Person of Interest from *B* is also an instance of Person of Interest of *A* or we have to explicitly define an assertion that says Persons of Interest of A subsumes Persons of Interest of B. Instead, the system should automatically make the inference without requiring the end-user to have any knowledge of the ontologies.

Inferring subsumption may not be enough to use instances of *B* in the query result for *A*. For the example we gave, it may be sufficient because the instances of *B* have the same attributes as instances of *A*. What if *A* subsumes *B*, but the name of the attributes aren't the same? Suppose that the name of the date of birth attribute is birth_date in *A*, but date_of_birth in *B*. Even if it can be inferred that *A* subsumes *B*, it doesn't solve the problem of converting instances of *B* to instances of *A*. The data space would also need to execute logic that performs the conversions. In programming languages, this logic is called a *coercion*. We consider the process of inferring subsumption and executing coercions a method of resolving *semantic heterogeneity* automatically.

In the context of resolving semantic heterogeneity in the Semantic Web, we can think of classes as schemas for linked data. Each schema specifies the structure of objects. Semantic heterogeneity occurs when those two schemes define different structures, but have the same meaning. When a schema subsumes another schema, the meaning of the schemas are close enough that instances of the subsumed schema can be treated as instances of the parent schema. Since the structures of the instances may be different, a coercion must be used so that the application expecting instances of the parent schema can use instances of the subsumed schema.

In this paper, we present an overview of a type system called TTIQ that can be used for modelling the Semantic Web. By using a type system little or no A-box statements will be needed. Some of the T-box statements that are used to specify subsumption and equality would not be needed. Having less work to maintain an ontology should make it easier to integrate different data sources that are developed and managed independently.

### 1.1 Related Work
The main difference between TTIQ and other Semantic Web type systems is the ability to automatically resolve *attribute name semantic heterogeneity*, different attribute names that have the same meaning. The type systems we are aware for the Semantic Web have the same semantics as the object-oriented languages [1] or less expressive than type system of TTIQ [2]. None of them define rules for relating types with different attribute names, a requirement for resolving attribute name semantic heterogeneity.

TTIQ is closely related to S-DTT [3] because TTIQ and S-DTT use *dependent types* for semantic modelling. In S-DTT, dependent product types are used to define binary relations or properties. In our system, we use *concrete data types* to define properties. By using concrete data types, we can represent triples symbolically as they would appear in a data store.

**Table 1: Subset of TTIQ Subtying Rules**

| $\dfrac{P \text{ is a primitive type}}{\Gamma \vdash P \leq \textbf{string}}$ (STR-PRIM) | $\dfrac{}{\Gamma \vdash T \leq T}$ (IDENTITY) | $\dfrac{\Gamma \vdash t \in T \quad \Gamma \vdash T \leq U}{\Gamma \vdash \exists x{:}T.T' \leq U}$ (∃-FIRST) |
|---|---|---|
| $\dfrac{c \preccurlyeq c' \quad \Gamma \vdash T_1 \leq T'_1 \quad \cdots \quad \Gamma \vdash T_n \leq T'_n}{\Gamma \vdash c \textbf{ of } T_1 \times \cdots \times T_n \to C \leq c' \textbf{ of } T'_1 \times \cdots \times T'_n \to C'}$ (CONCRETE) | | $\dfrac{m \leq n \quad \Gamma \vdash T_1 \leq T'_1 \quad \cdots \quad \Gamma \vdash T_m \leq T'_m}{\Gamma \vdash a_1{:}T_1 \times \cdots \times a_n{:}T_n \leq a_1{:}T'_1 \times \cdots \times a_m{:}T'_m}$ (SYN-REC) |
| $\dfrac{m \leq n \quad a_1 \preccurlyeq a'_1 \quad \cdots \quad a_m \preccurlyeq a'_m \quad \Gamma \vdash T_1 \leq T'_1 \quad \cdots \quad \Gamma \vdash T_m \leq T'_m}{\Gamma \vdash a_1{:}T_1 \times \cdots \times a_n{:}T_n \leq a'_1{:}T'_1 \times \cdots \times a'_m{:}T'_m}$ (SEM-REC) | | |

## 2. SEMANTIC MODELING WITH TTIQ

In this section, we demonstrate how a type system can be used to model the Semantic Web. In particular, we show how to represent linked data along with its semantics in a type system, how heterogeneity is resolved automatically using coercions, and how querying is performed by proving *type judgments*. We assume that the majority of our readers do not conduct research in type systems. Therefore, we omit some of the rigor found in a definition of a type system. Instead, we introduce portions of TTIQ as we present examples so that we do not lose the interest of readers.

### 2.1 Representing linked data

We start by demonstrating how a triple is represented using TTIQ. A triple is defined in TTIQ as a term of the form $p(s,o)$ where $p$ is a concrete datatype constructor, $s$ is a term, and $o$ is a term. A concrete datatype constructor is an operator used to create members of a concrete datatype. Therefore, all triples created with the constructor belong to the same type. For example, if $x$, $y$, and $z$ are instances of the class Human and hasSibling is a type constructor, then $hasChild(x, y)$ and $hasChild(x, z)$ are each objects that belong to the type hasChild.

We use this approach so that we can model object properties as types. More specifically, for any object property $p$, we model it in TTIQ as the type $p \textbf{ of } T \times T'$ where $T$ is the type representing the domain and $T'$ is the type representing the range.

In TTIQ, we model all essential properties as *fields* of a *record type*, even if the essential property is an object property. Non-essential properties use concrete datatypes and fields. If the property can be added after an instance is created, then we usually represent it as a concrete datatype.

We use records to model instances of classes. A class is defined as a type that is *inhabited* by records. The simplest types that contain records are the *record types*. If $a_1, \dots, a_n$ are labels and $T_1, \dots, T_n$ are types then $a_1{:}T_1 \times \cdots \times a_n{:}T_n$ is a record type. We can use a record type to model Person of Interest from the introduction as name:**string** × date_of_birth:**date**. The types **string** and **date** are the primitive types of TTIQ for strings and dates, respectively.

Similar to description logic, we can define classes as axioms. In TTIQ, we define the axioms using the *quantified type constructors* $\forall x{:}T.T'$ and $\exists x{:}T.T'$. The *dependent product type*, $\forall x{:}T.T'$, is read as for all $x$ with type $T$ such that $T'$." The *dependent sum type*, $\exists x{:}T.T'$, is read as there exists an $x$ of type $T$ such that $T'$.

Here's an example to demonstrate how we use dependent types to represent classes. Suppose we represent the class Human as the record type with one field gender:{'Male', 'Female', 'Unknown'}. The type {'Male', 'Female', 'Unknown'} is an example of an enumerated type in TTIQ. In TTIQ, if $s_1, \dots, s_n$ are strings, then $\{s_1, \dots, s_n\}$ is an enumerated type. We can define Man as

$$\exists x{:}\text{Human}. x \cdot \text{gender} \equiv \text{'Male'} \quad (1)$$

The term $x \cdot$ gender in (1) is an instance of *record selection*. In other words, if $t$ is a term that *evaluates* to $(a_1 = t_1, \dots, a_i = t', \dots, a_n = t_n)$ and $a$ is equal to $a_i$, then $t \cdot a$ evaluates to $t'$. The term $x \uparrow \text{gender} \equiv \text{'Male'}$ is concrete syntax for the term $\equiv (x.\text{gender},\text{'Male'})$ where $\equiv$ is a predicate symbol representing string equality. In TTIQ, we assume the existence of a set of predicate symbols and function symbols. More formally, given a predicate symbol $P$ and terms $t_1, \dots, t_n$, $P(t_1, \dots, t_n)$ is a term. Likewise given a function symbol $f$ and terms $t_1, \dots, t_n$, $f(t_1, \dots, t_n)$ is a term.

In TTIQ, we only define computation on a few terms and rely on an external interpreter to define computation on others. In particular, in TTIQ we use an interpreter to define evaluation of the application of function symbols and predicate symbols, that is terms of the form $f(t_1, \dots, t_n)$ and $P(t_1, \dots, t_n)$. To use TTIQ for the Semantic Web, using an interpreter is essential. If we didn't use an interpreter, we would need to define a means on how to define the semantics of each function symbol and predicate symbol operationally in TTIQ. In other words, we would have to include a general purpose programming language.

Here's an example to demonstrate the difference between using TTIQ for semantic modelling compared to description logic. In description logic, we define a Mother as Woman ⊓ ∃hasChild.Person. This formula says a mother is a woman that has a person as a child. In other words, in order for a woman $w$ to be a mother there must be a triple hasChild$(w, p)$ where $p$ is a Person. In TTIQ, we write this as

$\exists w.\text{Woman}. \exists r{:}\text{hasChild}. \exists p{:}\text{Person}. r = \text{hasChild}(w, p) \quad (2)$

In TTIQ, this type is inhabited by terms of the form $\bigl(w, (r, (p, \textbf{true}))\bigr)$. We should be able to conclude that every mother is a woman. However, a woman is just $w$ and a mother is $\bigl(w, (r, (p, \textbf{true}))\bigr)$. In order to deduce that every mother is a woman, we need to use *coercions*.

A coercion is a mapping between types that preserves the order of the elements of the types. We can derive a coercion from $T$ to $T'$ if $T$ is a subtype of $T'$. In description logic, we do not need coercions because if $T$ subsumes $T'$ then every instance of $T$ is an instance of $T'$. In type systems, subsumption means if $T$ is a subtype of $T'$, then every term $t$ that has type $T$ also has type $T'$. In TTIQ, this property does not hold. However, we can conclude that if $T$ is a subtype of $T'$, then there is a coercion $\kappa$ from $T$ to $T'$ such that for any term t that has type $T$, $\kappa(t)$ has type $T'$.

In TTIQ, Mother is a subtype of Woman. The coercion between Mother and Woman takes the pair $\bigl(w, (r, (p, \textbf{true}))\bigr)$ and returns $w$. We describe coercions in more details in Section 2.3.

**Table 2: Coercion construction from subtyping rules.**

| | |
|---|---|
| (STR-PRIM) | If $P \neq$ **num**, then $\rho: P \hookrightarrow$ **string** is the coercion where for each $t \in P$, $\rho(t) = \text{'}t\text{'}$ |
| | If $P =$ **num**, then $\gamma_k:$ **num** $\hookrightarrow$ **string** is the coercion where for each $n \in P$, $\rho_k(n) = \text{'}00..00n\text{'}$. The number of leading zeroes plus the number of characters for $n$ is equal to $k$. We require the leading zeroes so that $\gamma(12) \leq \gamma(2)$. |
| (IDENTITY) | The identity coercion $\iota: T \hookrightarrow T$. Given any $t \in T$, $\iota(t) = t$. |
| (SYN-REC) | Let $\kappa_i: T_i \hookrightarrow T_i'$ be the coercion extracted from $\Gamma \vdash T_i \leq T_i'$. Then $\kappa: a_1: T_1 \times \cdots \times a_n: T_n \hookrightarrow a_1: T_1' \times \cdots \times a_m: T_m'$ is the coercion where $\kappa(a_1 = t_1, \ldots, a_n = t_n) = (a_1 = \kappa_1(t_1), \ldots, a_n = \kappa_n(t_n))$ |
| (SEM-REC) | |
| (∃-FIRST) | Let $\kappa': U \hookrightarrow T$ be the coercion extracted from $\Gamma \vdash T \leq U$. Then $\kappa: T \hookrightarrow U$ is the coercion where $\kappa(t, p) = t$ |
| (CONCRETE) | Let $\kappa_i: T_i \hookrightarrow T_i'$ be the coercion extracted from $\Gamma \vdash T_i \leq T_i'$. Then $\kappa: c$ **of** $T_1 \times \cdots \times T_n \to C \to c'$ **of** $T_1' \times \cdots \times T_n' \to C'$ is the coercion where $\kappa(c(t_1, \ldots, t_n)) = c'(\kappa_1(t_1), \ldots, \kappa_n(t_n))$ |

## 2.2 Subtyping

Given a type $T$ and $T'$ we write $T$ is a subtype of $T'$ as $T \leq T'$. To determine whether $T \leq T'$, we use a *subtype judgment*. A subtype judgment is written as $\Gamma \vdash T \leq T'$. $\Gamma$ represents a sequence of declarations and is called an *environment*. A declaration is written as $x: U$ where $U$ is a type and $x$ is a variable, function symbol, predicate symbol, or concrete datatype constructor. If $x$ is a variable, then $U$ can be any type. If $x$ is a function symbol, then $U$ is the type of the form $T_1 \times \cdots \times T_n \to P$ where each $T_i$ is any type and $P$ is a primitive type. If $x$ is a predicate symbol, then $U$ is the type of the form $T_1 \times \cdots \times T_n \to$ **bool**. If $x$ is a type constructor for a concrete datatype, then $U$ is the type of the form $T_1 \times \cdots \times T_n \to C$ where $C$ is the name of a concrete datatype. In practice, we expect $c$ and $C$ to be the same.

To determine whether $\Gamma \vdash T \leq T'$ is valid, we have to prove it. Proofs of subtype judgments are constructed using inference rules. We define a subset of the inference rules in the Table 1. The rule STR-PRIM says that every primitive type is a subtype of **string**. The other primitive types are **num**, **bool**, **uri,** and **date**. Therefore, **num** $\leq$ **string**, **bool** $\leq$ **string**, **date** $\leq$ **string**, **uri** $\leq$ **string**, and **bool** $\leq$ **num**. The SYN-REC rule states that if a class is defined as a record type then it subsumes another class if it has the same fields as the parent class. In practice, this may be too restrictive. Therefore, we also define subtyping on record types to allow the labels to be syntactically different, but semantically similar. To determine semantic similarity we assume the existence of a taxonomy of labels. We model the taxonomy as a partial ordering of labels. We denote the partial order as $\preccurlyeq$. The SEM-REC rule allows us to prove that

$$\text{name}:\textbf{string} \times \text{date\_of\_birth}:\textbf{date} \times \text{ethnicity}:\textbf{string} \quad (3)$$

is a subtype of

$$\text{name}:\textbf{string} \times \text{date\_of\_birth}:\textbf{date} \quad (4)$$

If the semantic network of attribute names specified that birth_date $\preccurlyeq$ date_of_birth, then

$$\text{name}:\textbf{string} \times \text{birth\_date}:\textbf{date} \times \text{ethnicity}:\textbf{string} \quad (5)$$

is also a subtype of (4). If we defined (5) as

$$\text{ethnicity}:\textbf{string} \times \text{birth\_date}:\textbf{date} \times \text{name}:\textbf{string} \quad (6)$$

then (6) isn't a subtype of (4) according to our definition because name isn't the first field. We can resolve this by defining subtyping to ignore the ordering of fields. For brevity, we do not formally define it here, but assume that our definition ignores the order of attributes. So (6) is a subtype of (4) in TTIQ.

The rule ∃-FIRST allows us to prove that Mother is a subtype of Woman. The left antecedent of ∃-FIRST contains a *type judgment*. The type judgment $\Gamma \vdash t \in T$ is used to assert that the term $t$ is a member of the type $T$ or $t$ inhabits the type $T$. Due to space limitations, we omit presenting the type judgment rules. Please see [4] for a subset of the type judgment rules of TTIQ. The antecedent on the right of ∃-FIRST means that in order to prove that Mother is a subtype of Woman we only need to prove that $\Gamma \vdash$ Woman $\leq$ Woman. This judgment is proven by the IDENTITY rule.

The CONCRETE rule allows us to define relationships between object properties. The rule says that in order for two properties to be related, the names of the type constructors have to be semantically related and the input types and output types must be related. The semantic relationship is modelled as a partial ordering on type constructor names.

From the proof of a subtype judgment, we can extract a coercion that maps terms of the subtype to terms in the supertype. In the next section, we describe how coercions are extracted from proofs.

## 2.3 Coercions

A coercion is a mapping between two types that preserve the order of terms. We denote the coercion $\kappa$ between $T$ and $T'$ as $\kappa: T \hookrightarrow T'$. If $\vdash t \in T$ and $\kappa(t) = t'$, then $\vdash t' \in T'$. Furthermore, if $\ll_T$ is the *natural partial ordering* on $T$ and $\ll_{T'}$ is the natural partial ordering on $T'$ and if $\kappa(t_0) = t_0'$ and $\kappa(t_1) = t_1'$, then if $t_0 \ll_T t_1$ then $t_0' \ll t_1'$. In TTIQ, every type has a natural partial ordering. For example, the natural partial ordering on **num** is $\leq$. The natural partial ordering on **string** is the lexigraphical ordering based on the ordering of Unicode characters. From these natural partial orderings, we can define the natural partial orderings for other types inductively. For brevity, we omit defining the natural partial orderings of the remaining types.

In TTIQ, if $\Gamma \vdash T \leq T'$ then there exists a coercion $\kappa: T \hookrightarrow T'$. We prove this informally by showing how coercions are constructed for different subtype judgments. In particular, for each rule in Table 1, we describe how to create the corresponding coercion. Table 2 contains rules for constructing coercions.

## 2.4 Resolving semantic heterogeneity using coercions

In the example from Section 1, an application uses a query to retrieve Persons of Interests from a data space. In between the application and the data sources of the data space is a middle tier that implements TTIQ. Whenever a contributor wants to include its data in the system, it registers with the middle-tier by supplying ti with types representing its schema. When the middle tier receives the type, say $T$, it determines which other types it is related to it by proving judgments of the form $\Gamma \vdash T \leq T'$ and $\Gamma \vdash T' \leq T$.

Recall from Section 1, we referred to the Persons of Interest from the news organization as $A$. Let's assume that an application makes the following request to retrieve Persons of Interest.

$$\text{SELECT ?x FROM ?x a } A \qquad (6)$$

This query will return terms of type $A$. Suppose that two new sources of Persons of Interest are added that have the types in (3) and (5). Let's call these types $B$ and $C$, respectively. The middle tier will deduce that $\Gamma \vdash B \leq A$ and $\Gamma \vdash C \leq A$. It will also construct the coercion $\kappa_0: B \hookrightarrow A$ where $\kappa_0((\text{name} = s, \text{date\_of\_birth} = d, \text{ethnicity} = e)) = (\text{name} = s, \text{date\_of\_birth} = d)$. And it will derive the coercion $\kappa_1: C \hookrightarrow A$ where $\kappa_1((\text{name} = s, \text{birth\_date} = d, \text{ethnicity} = e)) = (\text{name} = s, \text{date\_of\_birth} = d)$. When the query in (6) is executed after the new data source is added, the query engine determines which data sources to request for data by determining which data sources have types that are subsumed by $A$. By subsumed, we mean that there exists a coercion from the data source's type to $A$. Therefore, in our examples, requests are sent to the data sources that contain $A$, $B$, and $C$. When the query engine gets terms from the data sources for $B$ and $C$, it uses the coercions $\kappa_0$ and $\kappa_1$ to convert the terms to instances of $A$. By using this approach, the application is able to use new data sources without being changed. Also, the data contributors do not have to modify their systems to be used by the application.

## 2.5 Queries as Type judgments

Executing a query in TTIQ is performed by finding a proof of a type judgment. In TTIQ a type judgment has the form $\Gamma \vdash t \in T$. For queries, the type judgments have the form

$$\Gamma \vdash \exists x: T. T' \in \mathbf{prop} \qquad (7)$$

and

$$\Gamma \vdash \exists X \leq T. T' \in \mathbf{prop}. \qquad (8)$$

.

The proof of (7) only results in one term being found. However, the one term could be a type. The proof of (8) requires finding a collection of terms. The judgment in (8) contains a *bounded quantifier*. A bounded quantifier restricts instances of $X$ to be subtypes of $T$. This means that to prove this judgment a subtype of $T$ has to be found that satisfies $T'$.

The terms needed to prove (7) and (8) are TTIQ representations of data in a data space. The data space could be implemented using the TTIQ term language or it could use some other language, such as RDF(S). The only requirement is that the data space returns TTIQ terms to the TTIQ middleware. This means the terms could be created from an analytic or even by a human when alerted by the query engine. We believe an open-ended, but restricted, approach to query answering is a must-have in an open data space such as the Semantic Web.

In TTIQ all proofs are formal. In other words, the execution process of a query can only use specific steps defined by TTIQ. One of these steps allow an analytic or an external procedure to be invoked. Each of these procedures must formally specify their interfaces. In other words, each analytic and procedure must specify its precondition and postcondition as a type. The type has the form $\forall x: T. \phi \to \exists y: U. \psi$. This type says that the input to the procedure must be a term of type $T$ and the input must satisfy the condition defined by the type $\phi$. The procedure will produce a term of type $U$ that satisfies $\psi$. This is the shorter version of an interface. The form of the interface that will most likely be used is the type in (9).

$$\forall X \leq U. \forall x: T. (\phi \to \exists Y \leq U'. \forall y: T'. \psi) \qquad (9)$$

The type in (8) means that the input to the procedure could be a class $U$ and the output could be the class $U'$. This means that procedures could use knowledge discovery or machine learning algorithms. As a result, we can include analytics to generate data to answer queries.

## 2.6 Undecidability

There could be multiple proofs of $\Gamma \vdash A \leq B$. Therefore, there could be different coercions. Since proving $\Gamma \vdash A \leq B$ is performed when a data source is added, we can make proving it part of a tool for adding data sources. Then the person adding the data source could select the coercions to use.

TTIQ uses higher-order types which can make proving type judgments undecidable. This means querying is undecidable. We can overcome undecidability by defining a query process that requires a human user to guide it. We refer the reader to our approach on *interactive querying* in [4] as a possible solution to undecidability.

TTIQ doesn't define a mechanism for managing the taxonomy of labels and constructor names. Instead, other approaches that address ontology merging and matching could be used. For example, we could incorporate a tool such as Agreement Maker [5] to manage merging taxonomies.

## 3. CONCLUSION

We presented a type system, called TTIQ, to represent knowledge. Using TTIQ, we demonstrate how to use type constructors as predicates in order to create linked data. We showed how subtyping can be used to determine whether heterogeneity could be resolve automatically using coercions. We described how querying can be defined before by proving that a data space contains terms that belong to a type and satisfy specific conditions.